\documentclass[
 reprint,
 floatfix,
 amsmath,amssymb,
 aps,
]{revtex4-2}
\usepackage{amsmath}
\usepackage{graphicx}
\usepackage{dcolumn}
\usepackage{bm}
\usepackage{braket}
\usepackage{}
\usepackage{enumitem}

\begin{document}

\title{Emerging Nonlocal K\"{a}ll\`{e}n-Lehmann 
Higgs Spectra at the LHC 
}

\author{Stathes Paganis}
\thanks{paganis@phys.ntu.edu.tw}
\affiliation{
 Department of Physics, National Taiwan University, 
No 1, Sec 4, Roosevelt Road, Taipei 10617, Taiwan,\\
Leung Center for Cosmology and Particle Astrophysics,
National Taiwan University, Taipei 10617, Taiwan
}

%

%
\date{\today}

\begin{abstract}
Electroweak symmetry breaking may arise from emergent nonlocal K\"{a}ll\`{e}n-Lehmann spectral densities in Hamiltonians with multiscalar interactions. The nonlocality scale $\Lambda_{NL}$ emerges naturally from the exponentially increasing degeneracy of mass eigenstates in the Higgs two-point function at scales $p^2 \geq \Lambda^2_{NL}$. Following the renormalization of the nonlocal Higgs propagator, we provide a framework for deriving analytic expressions for non-perturbative scattering amplitudes. We demonstrate that for energies exceeding the nonlocality scale, scattering amplitudes are exponentially suppressed. Furthermore, the real part of the Higgs self-energy is suppressed at deep spacelike momenta ($p^2 \sim -\Lambda^2_{NL}$), offering a solution to the Hierarchy problem. Such nonlocal scalar sectors are accessible to current and future LHC runs. We argue that the nonlocal K\"{a}ll\`{e}n-Lehmann spectral density can be constrained through a simultaneous global fit of LHC measurements in exclusive channels, including di-Higgs, electroweak di-boson, and di-photon production. This approach represents a paradigm shift in the search for new physics at high-energy colliders.
\end{abstract}

\maketitle


\section{\label{intro}Introduction}
 
Experiments at the energy frontier in the Large Hadron Collider (LHC) have been performing measurements of scattering amplitudes with the goal to detect departures from the standard model (SM) of particle physics. These departures are typically attributed to the presence of new particles or interactions that can appear in final states as bound states, resonances, continua, or anomalous couplings. Treating SM as an effective theory and motivated by the hierarchy problem \cite{Weinberg:1975gm}, we expect that new interactions and particles will manifest themselves at some scale $\Lambda\sim\mathcal{O}$(TeV).

Several extensions beyond the Standard Model (BSM) introduce new physics and predict the presence of new states
at $\Lambda_{BSM}$.
These new-physics sectors are assumed to couple to SM particles. 
Examples include large extra spatial dimension models, such as those proposed by ADD \cite{Arkani-Hamed:1998jmv} and RS \cite{Randall:1999ee}, where Kaluza-Klein (KK) graviton modes couple to the SM.
In other frameworks \cite{Giudice:2016yja}, an infinite tower of massive spin-2 graviton KK modes is predicted. 
Of particular interest are non-gravitational models, including composite Higgs scenarios where heavy-quark bound states are predicted \cite{Contino:2010rs}. 
There is a class of models of BSM interactions
inspired by string-field theory that lead to violations of locality. 
The characteristic feature of these nonlocalizable theories is that the spectral density
$\rho_{BSM}(m^2)$ is not polynomially bounded, violating locality at some specific nonlocality scale.
Such structures 
appear in UV completions like string theory at the perturbative level \cite{deRham:2010kj,deRham:2014zqa,Cheung:2014dqa,Nortier:2021six}, through non-perturbative effects in the classicalization proposal \cite{Cachazo:2014xea}, and in Galileon theories \cite{Hinterbichler:2015pqa}. It has been conjectured that the UV description of Galileons—and by extension, theories of massive gravity, bi-gravity, and multi-gravity—violates the condition of polynomial boundedness while respecting unitarity, analyticity, and Lorentz invariance \cite{Galileon2015}.

In this work, we extend the SM Hamiltonian $\mathcal{H}_{0}$ by introducing a set of non-perturbative BSM sectors, $\mathcal{H}_{BSM}$
\begin{equation}
\mathcal{H}= \mathcal{H}_{0} + \mathcal{H}_{BSM}, \nonumber
\label{eq:H}
\end{equation}
which are characterized by an infinite series of multi-scalar local interactions. We argue that these sectors generate nonlocalizable spectral densities, thereby necessitating a nonlocal QFT framework. In parallel, we propose a general approach for extracting information about these sectors from scattering amplitudes by directly measuring the positive-definite spectral density of the BSM Hamiltonian.
The BSM sectors may consist of one or more new heavy scalars that are strongly coupled to the Higgs boson $h$. For example, a theory with a single extra heavy scalar $\phi$ can be modeled as follows:
\begin{equation}
\mathcal{H}_{int}=\lambda_3 h^2\phi + \lambda_4 h^3\phi + \dots
+\lambda_{k}h^{k-1}\phi + \dots ,
\label{eq:HINT}
\end{equation}
where the couplings $\lambda_{k}$ are of order unity. The central object in our proposal is the spectral density for a specific operator $\mathcal{O}$ in the Hamiltonian, defined as:
\begin{equation}
    \rho^{}_{\mathcal{O}}(p^2)=(2\pi)^3\sum_n \langle 0 |\mathcal{O}|n\rangle\langle n |{\mathcal{O}^\dagger}|0\rangle
     \delta^{(4)}(p-p_n),
    \label{rhodefinition1}
\end{equation}
where $\{|n\rangle\}$ represents a complete set of on-shell states. The term $|\langle 0|\mathcal{O}|n\rangle|^2$ denotes the probability that the operator $\mathcal{O}$ produces a state $n$ with momentum $p_n$. These states $|n\rangle$ can include both bound states and $n$-particle continua, yielding:
\begin{equation}
 \rho(p^2) = \sum_{i}Z_i \delta(p^2-m_i^2) + \sigma(p^2)\Theta(p^2-m^2_{th}), 
\label{eq:rho1} 
\end{equation}
where the residue $Z_i$ is the probability for the field operator to produce the bound state $i$ from the vacuum, and $\sigma(p^2)$ represents the contribution from the continuum (or multiple continua). For the continuum description, the sum in Eq.~\ref{rhodefinition1} is replaced by a sum over integrals over the Lorentz Invariant Phase Space (LIPS).

In BSM theories such as those described by Eq.~\ref{eq:HINT}, the continuum thresholds are crossed as the energy scale $p^2 \gg m_h^2$ increases. In this regime, the Higgs two-point function accounts for $n$-loop on-shell exchanges of $n+1$ scalars, where $n$ is determined by the kinematically available phase space. Because these exchanges are dominated by the lightest scalar (the Higgs boson $h$), the mass eigenstates of the Hamiltonian exhibit a high degree of density. As we will demonstrate, for each additional on-shell Higgs exchange, the density of states $\rho(s)$ grows exponentially with the square of the center-of-mass energy $s=p^2$.
The spectral density in a local theory is polynomially bounded; exponentially growing densities—where $\rho(p^2) \sim \exp(p^2/\Lambda^2)$—are strictly nonlocalizable. Remarkably, although the interactions in the Hamiltonian of Eq.~\ref{eq:HINT} are individually local, the resulting spectral density can become nonlocalizable. This leads to propagators that exhibit the same behavior as scalar propagators in infinite-derivative nonlocal Quantum Field Theories (NLQFT). Consequently, the massive scalar renormalized propagator takes the following form:
\begin{equation}
\widetilde{G}\left(p^2\right) =
\frac{e^{p^2/\Lambda_{NL}^2}}
{p^2-m^2_{h}-ie^{p^2/\Lambda_{NL}^2}\textrm{Im}[\Sigma(p^2)]+i\epsilon}.
\end{equation}
The nonlocality scale $\Lambda_{NL}$ represents the energy scale at which nonlocal effects become dominant, serving as a natural cutoff that renders the theory UV finite. Since self-energy also exhibits an 
$\exp(p^2/\Lambda_{NL}^2)$ dependence, the resummed renormalized propagator for $p^2 > \Lambda_{NL}^2 \gg m_h^2$ becomes exponentially suppressed as a function of $p^2$. A crucial aspect of this NLQFT description is that the free propagator of the theory, in the absence of BSM interactions, remains the standard SM propagator. Nonlocal effects, manifested through the $\exp(p^2/\Lambda_{NL}^2)$ operator, emerge only when on-shell multi-scalar exchanges are activated.
A compelling connection between multiparticle interactions and nonlocality was established in \cite{PhysRevD.104.015028}, where it is argued that asymptotically nonlocal field theories interpolate between Lee-Wick theories with multiple propagator poles and ghost-free nonlocal theories. Similarly, \textit{Higgsplosion} models incorporate multiparticle interactions that could lead to nonlocalizable spectral densities \cite{Khoze:2017ifq}, further motivating the framework proposed here.

The presence of BSM sectors with nonlocalizable spectral densities is expected to yield observable effects at the LHC. In this work, we argue that the most general approach to probe for new physics is the direct measurement of the spectral density $\rho(m^2)$. This is achieved by performing a combined fit across several final states governed by the same BSM operators within the interaction Hamiltonian, $\mathcal{H}_{int}$. Candidate final states for such an analysis include di-Higgs ($hh$), electroweak di-boson ($VV$), $Vh$, and di-photon ($\gamma\gamma$) production.

This paper is organized as follows: In Section~\ref{KLSD}, we introduce the Higgs spectral densities. Section~\ref{NLQFTSD} provides a discussion of nonlocal QFTs and nonlocalizable densities. In Section~\ref{RENORM}, we detail the renormalization of the nonlocal Higgs propagator. Finally, in Section~\ref{PHENO}, we propose potential experimental signals that are sensitive to the nonlocal modifications of SM interactions.

%
%
\section{\label{KLSD}K\"{a}ll\`{e}n-Lehmann Higgs spectral density}

The two-point correlation function for an external 
momentum $p$, expressed in terms of operators in the 
SM interaction Lagrangian, is given by:
\begin{equation}
\Delta(p^2) =\langle \mathcal{O}_{SM}(p)\mathcal{O}_{SM}(-p) \rangle. \nonumber
\end{equation} 
A general expression for a scalar two-point function is an integral representation in terms of a linear combination of Feynman propagators, weighted by a real spectral density $\rho(m^2)$. This is the K\"{a}ll\`{e}n-Lehmann (KL) spectral representation \cite{Kallen:1949bza,Lehmann:1954xi}. In momentum space, the Higgs Green's function $G_h(p^2)$ is expressed via the KL representation as:
\begin{eqnarray}
G_h(p^2)
&=&-i\Delta_h\left(p^2\right)
= \int_{0}^{\infty} \frac{\rho^{}_h\left(m^2\right)}{p^2-m^2+i\epsilon} 
dm^2,
\label{KL1}
\end{eqnarray}
where the Higgs spectral density follows from Eq.~\ref{eq:rho1}:
\begin{equation}
\rho^{}_h(m^2)=Z\delta(m^2-m_h^2)+\sigma^{}_{h}(m^2)\Theta(m^2-m_{th}^2).
\end{equation}
The delta function represents a bound state at the physical Higgs mass $m_h$, while $m_{th}^2$ denotes the threshold of the lowest-energy continuum. Consequently, $\sigma_h$ describes the multi-particle continua—specifically the two-Higgs states for $p^2 > 4m_h^2$ and the three-Higgs states for $p^2 > 9m_h^2$—generated by the trilinear $h^3$ and quartic $h^4$ SM operators, respectively. The Higgs propagator then takes the form:
\begin{eqnarray}
G_h(p^2)
&=&
\frac{Z}{p^2-m_h^2+i\epsilon} 
+
\int_{4m_{h}^2}^{\infty} \frac{\sigma^{}_{h}(m^2)}{p^2-m^2+i\epsilon}dm^2. 
\label{eq:KL1}        
\end{eqnarray}

The KL representation provides the most general framework to express the propagation of a scalar $h$ in the presence of interactions without requiring perturbativity. Following \cite{Englert:2019zmt,Banks:2020gpu}, if we switch on new BSM interactions at a high-energy scale $m^2_{th}$, the resulting Higgs spectral density $\rho_h(m^2)$ can be written as:
\begin{equation}
    \rho^{}_h(m^2) = \rho^{}_{SM}(m^2) + \rho^{}_{BSM}(m^2),
\end{equation}
subject to the condition:
$$\rho^{}_{BSM}(m^2 < m_{th}^2)=0.$$ 
This allows us to decouple the modifications to the Higgs self-energy arising from BSM interactions from those of the SM. We make the strong assumption that the BSM sector manifests at a scale $p^2 = m_{th}^2 \simeq m_{\phi}^2$, where $m_{th}^2 \gg m_h^2$, representing the threshold of a new heavy scalar sector. Consequently, the Higgs propagator can be expressed as:
\begin{eqnarray}
G_h(p^2)
&=&
\frac{1}{p^2-m_h^2-i\textrm{Im}[\Sigma_{SM}(p^2)]+i\epsilon} \nonumber\\
&+&
\int_{m_{th}^2}^{\infty} \frac{\rho^{}_{BSM}(m^2)}{p^2-m^2+i\epsilon}  dm^2 ,
\label{eq:KL2}        
\end{eqnarray}
where the first term is the renormalized Higgs propagator, including the imaginary contributions to the self-energy from the SM continua described in Eq.~\ref{eq:KL1}.

\section{\label{NLQFTSD}Nonlocal QFT and Nonlocalizable spectral densities}

Nonlocal Quantum Field Theories (NLQFTs) \cite{Tomboulis:1997gg,Modesto:2011kw,Ghoshal:2017egr,Buoninfante:2018mre,Ghoshal:2020lfd,Nortier:2023dkq,Chattopadhyay:2023nbj,PinedoSoto:2025hel} employ entire transcendental functions of higher-derivative operators, $\mathcal{F}(\Box/\Lambda^2)$, to regularize the ultraviolet (UV) behavior of loop amplitudes without violating the fundamental requirements of unitarity and causality. By ensuring that the propagator possesses no additional poles, NLQFTs remain super-renormalizable or even finite, effectively shielding the Higgs potential from the large radiative corrections that characterize local theories.
It has been argued that such theories may resolve the Higgs mass fine-tuning problem by replacing cutoff-dependent UV divergences with regulated loops, thereby eliminating quadratic divergences \cite{Abu-Ajamieh:2023syy}. In this section, we briefly introduce infinite-derivative nonlocal scalar actions and discuss the conditions under which nonlocalizable spectral densities emerge.

\subsection{Infinite Derivative Nonlocal Scalar Action}

A general NLQFT action for a single scalar field $\phi(x)$ is given by \cite{Buoninfante:2018mre}:
\begin{equation}
\mathcal{A} = \int d^4x\left[-\frac{1}{2}\phi \mathcal{F}(\Box) \left(\Box+m_{\phi}^2\right) \phi - V\left(\phi\right)\right],
\label{NLagrangian}
\end{equation}
where $\Box=\partial_{\mu}\partial^{\mu}$ and the operator $\mathcal{F}(\Box)$ represents an infinite series of derivatives. Selecting $\mathcal{F}(\Box)=\exp(\Box/\Lambda_{NL}^2)$ yields an action of the form:
\begin{equation}
\mathcal{A} = \int d^4x\left[-\frac{1}{2}\phi e^{\frac{\Box}{\Lambda_{NL}^2}} \left(\Box+m_{\phi}^2\right) \phi - V\left(\phi\right)\right].
\label{NLagrangian2}
\end{equation}
Employing a $(+,-,-,-)$ metric signature where $p^2$ corresponds to the Minkowski momentum squared (such that $p^2 = -\Box$ in the operator sense), the propagator is given by \cite{Buoninfante:2018mre}:
\begin{equation}
\Delta\left(p^2\right) =
\frac{e^{p^2/\Lambda_{NL}^2}}{p^2-m^2_{\phi}+i\epsilon}.
\label{prop1}
\end{equation}
Crucially, when the potential is switched off, the theory effectively recovers the standard Feynman propagator behavior in the limit $p^2/\Lambda^2_{NL}\rightarrow 0$. The presence of nonlocal operators $e^{\Box/\Lambda_{NL}^2}$ in the kinetic term is mathematically equivalent to preserving local kinetic terms while shifting the nonlocal operators into the potential. In this representation, interaction vertices appear effectively "smeared" at the nonlocality scale $\Lambda_{NL}$.
It has been proposed that such nonlocal terms could be detected at the LHC as modifications to SM scattering amplitudes via a form factor \cite{BISWAS2015113,Su:2021qvm}:
\begin{equation}
\frac{\sigma}{\sigma_{SM}} \simeq e^{\frac{p^2}{\Lambda_{NL}^2}}.
\label{NLxsection}
\end{equation}
Our objective is to provide a more precise treatment of this modification. We demonstrate that after the renormalization of the Higgs self-energy, the cross-section becomes strongly suppressed for $p^2 \gtrsim \Lambda^2_{NL}$. This behavior arises because the exponentially increasing imaginary part of the self-energy, upon Dyson resummation, dominates the propagator and suppresses the scattering amplitude. Consequently, the exponential departures predicted by Eq.~\ref{NLxsection} are only expected in the moderate energy regime where $p^2 \sim \Lambda^2_{NL}$. Current LHC data show no evidence of such anomalous cross-sections, placing lower limits on the nonlocality scale in the several TeV range~\cite{CMS:2018dqv,CMS:2022tqn,Su:2021qvm}.

\subsection{\label{sec:level2} Nonlocalizable Spectral Densities}

The growth of the spectral density $\rho(m^2)$ as $m^2 \to \infty$ determines the localizability of a theory. A theory is defined as localizable if $\rho(m^2)$ is bounded in the UV by a polynomial of finite order. More generally, we characterize a theory as \textit{strictly localizable}, \textit{quasi-local}, or \textit{nonlocalizable} by expressing the high-energy behavior of the spectral density as \cite{Galileon2015,Tomboulis:2015gfa,Buonin2022}:
\begin{equation}
\rho(m^2) \sim \exp(m^{2\alpha}) \times \text{subdominant terms}.
\label{rhoNLclassif}
\end{equation}
The degree of localizability is then determined by the index $\alpha$:
\begin{itemize}
\item $\alpha < 1/2$: The theory is strictly localizable (Jaffe localizable).
\item $\alpha = 1/2$: The theory is quasi-local (the boundary of localizability).
\item $\alpha > 1/2$: The theory is nonlocalizable.
\end{itemize}
In gravitational theories, the spectral densities of generic operators typically grow faster than linear exponentials; consequently, operator-valued fields in quantum gravity are fundamentally nonlocalizable \cite{Galileon2015}. For example, the density of states at high energies in gravity is expected to be dominated by black hole states, which scale according to the Bekenstein-Hawking entropy $S_{\text{BH}}$:
\begin{equation}
\rho(m^2) \sim \exp(S_{\text{BH}}) =
\exp\left[c\left(\frac{m}{M_{Pl}}\right)^{\frac{d-2}{d-3}}\right],
\end{equation}
where $d$ is the spacetime dimensionality. For $d \geq 4$, it follows that $\alpha \geq 1$, rendering the theory nonlocalizable. Similarly, Little String Theories \cite{Berkooz:1997cq,Seiberg:1997zk} exhibit the exponential growth characteristic of the Hagedorn density of states \cite{Galileon2015,Aharony:1998tt,Peet:1998wn,Minwalla:1999xi}. Specifically, the K\"{a}ll\'{e}n-Lehmann spectral density for the two-point function in these theories is expected to scale as:
\begin{equation}
\rho(m^2) \sim \exp\left(\frac{c M_{\text{string}} m}{\sqrt{N}}\right),
\end{equation}
where $M_{\text{string}}$ is the string scale and $N$ is the number of five-branes. Since this corresponds to $2\alpha = 1$ (or $\alpha = 1/2$), Little String Theories are classified as quasi-local.

\section{\label{RENORM} Renormalized K\"{a}ll\`{e}n-Lehmann spectral representation}

In this section, we renormalize the generalized K\"{a}ll\`{e}n-Lehmann spectral representation for nonlocalizable theories. Following the frameworks established in \cite{Englert:2019zmt} and \cite{Banks:2020gpu}, we identify the BSM physics with a new sector of scalar states in the interaction Hamiltonian that extends the SM (see Eq.~\ref{eq:H}). We generalize the interaction operators introduced in Eq.~\ref{eq:HINT} as:
\begin{equation}
    \lambda_n h\mathcal{O}_{BSM}= 
    \lambda_n 
    h^{k}\prod_{i=1}^{n-k}\phi_i  
    \hspace{0.5cm} n>k>0.
    \label{Lint2}
\end{equation}
While this expression accommodates multiple additional fields $\phi_i$, we restrict our current study to the case of a single heavy scalar $\phi$, as originally proposed in Eq.~\ref{eq:HINT}. For example, the lowest-dimension three-scalar interaction, $h^2\phi$, is recovered by setting $k=2$ and $n=3$. In the following, we first introduce the nonlocal K\"{a}ll\`{e}n-Lehmann spectral representation and subsequently discuss its renormalization.

\subsection{\label{sec3:nlkl} Nonlocal K\"{a}ll\`{e}n-Lehmann Spectral Representation and Self-Energy}

We begin by defining the two-point correlation function in momentum space for a BSM operator $\mathcal{O}_{BSM}$ with external momentum $p$:
\begin{equation}
\Delta(p^2) =\langle \mathcal{O}_{BSM}(p)\mathcal{O}_{BSM}(-p) \rangle. 
\end{equation}    
For the BSM operator $\mathcal{O}_{BSM}$ introduced in Eq.~\ref{Lint2}, the scalar propagator for the field $h$ is expressed as:
\begin{equation}
  G(p^2)= 
\frac{Z}{p^2-m_h^2+i\epsilon} 
+
\int_{m_{th}^2}^{\infty} \frac{\rho^{}_{\mathcal{O}^{}_{BSM}}\left(m^2\right)}{p^2-m^2+i\epsilon}  dm^2. 
\end{equation}
The spectral density $\rho^{}_{\mathcal{O}^{}_{BSM}}(m^2)$ constitutes a sum of individual spectral densities, each describing a continuum corresponding to a specific number of particle exchanges. The threshold $m_{th}^2$ is determined by the one-loop exchange of a Higgs boson $h$ and the scalar $\phi$, such that $m_{th}^2 = (m_h + m_{\phi})^2$. Within the continuum integral, a Heaviside step function $\Theta(m^2 - m^2_{th})$ is implicitly assumed, as in Eq.~\ref{eq:rho1}, ensuring that the spectral density vanishes below the physical threshold.


The Higgs self-energy is the fundamental building block of our framework. Defined as the sum of all one-particle irreducible (1PI) amputated two-point functions, its spectral representation is the following:
\begin{equation}
\Sigma_h(p^2) = -\int_{m^2_{th}}^{\infty} \frac{\rho_{\Sigma}^{}\left(m^2\right)}{p^2-m^2+i\epsilon} dm^2.
\label{eq:se1}
\end{equation}
For $p^2 > m_{th}^2$, the correlation function involves on-shell exchanges, and the integral corresponds to the imaginary part of the self-energy. Utilizing the Cauchy theorem \cite{Zwicky:2016lka} and assuming the integrand vanishes as $m^2 \rightarrow \infty$, we obtain the discontinuity relation:
\begin{equation}
\textrm{Im}[\Sigma\left( p^2 \right)] = \pi \rho_{\Sigma}^{}(p^2).
\label{eq:disc}
\end{equation}
The imaginary part is determined by the optical theorem as a sum over $n$-particle states:
\begin{equation}
2\mathrm{Im}[\Sigma(p^2)] = \sum_{n=1}^{\infty} \int d\Phi_n |\langle 0 |\mathcal{O}|n\rangle|^2,
\end{equation}
where the Lorentz-invariant phase space element $d\Phi_n$ is:
\begin{eqnarray}
 d\Phi_n &=& (2\pi)^4
 \delta^{(4)}\left(p-\sum_i^n q_i\right)
 \prod_i^n\frac{d^3p}{(2\pi)^32E}. \nonumber
\end{eqnarray}
For a BSM Hamiltonian with an infinite number of interaction terms, the 1PI spectral density $\rho_{\Sigma}(m^2)$ becomes:
\begin{eqnarray}
\rho^{}_\Sigma(m^2) &=& \rho^{}_{1L}(m^2) +  \rho^{}_{2L}(m^2) + \dots
+ \rho^{}_{L}(m^2) + \dots
\nonumber \\
&=& \frac{1}{2\pi}\sum_{n=1}^{\infty} \int d\Phi_n |\langle 0 |\mathcal{O}|n\rangle|^2,
\end{eqnarray}
where $L=n-1$ denotes the number of loops and $n$ the number of exchanged particles. A term with $n$ exchanges ($n-1$ loops) scales as:
\begin{eqnarray}
    \rho_{n-1} &\sim& (n!)^2 
    \left(\frac{1}{n!}
    \frac{\lambda^2}{\Lambda^{2(n-3)}}\right)
 \Phi_n
 = \frac{n!\lambda^2}{\Lambda^{2(n-3)}}\Phi_n.\nonumber
\end{eqnarray}
The phase space volume $\Phi_n$ for $n$ exchanges acquires factors of $1/(n-1)!$ due to the degeneracy of identical Higgs exchanges and energy conservation ($\sum E_i = \sqrt{s}$), scaling roughly as $\sim ((p^2-m^2_{th})/16\pi^2)^n$. This leads to the ratio:
\begin{equation}
\frac{\rho_{n+1}}{\rho_{n}} \simeq \frac{n!}{(n-1)!} \times \frac{1}{\Lambda^2} \frac{\Phi_{n+2}}{\Phi_{n+1}} \simeq \frac{1}{n-1} \frac{p^2-m^2_{th}}{16\pi^2\Lambda^2}.
\end{equation}
Summing these contributions results in a spectral density that grows exponentially with $m^2$, rendering the theory nonlocalizable:
\begin{equation}
\rho^{NL}_{\Sigma}(m^2) \simeq \exp\left(\frac{m^2-m_{th}^2}{\Lambda_{NL}^2}\right) \rho^{}_{1L}(m^2).
\end{equation}
Nonlocalizable K\"{a}ll\`{e}n-Lehmann representations were recently introduced in \cite{Paganis:2024end,Briscese:2024tvc} and proven in \cite{Briscese:2024tvc}. The nonlocalizable version of the full propagator is then:
\begin{eqnarray}
G(p^2)
&=& 
\frac{Z}{p^2-m_h^2+i\epsilon}+
\int_{m^2_{th}}^{\infty} 
\frac{\rho^{NL}_{}\left(m^2\right)}{p^2-m^2+i\epsilon} 
dm^2,
\label{nrKL}        
\end{eqnarray}
where $\rho^{NL}$ represents the full spectral density obtained after Dyson resummation.

\subsection{\label{Dyson} Self Energy: N-subtraction Renormalization}

Equation~\ref{eq:disc} allows us to determine the Higgs self-energy provided an expression for the total spectral density is known. However, proving this relation requires performing the contour integral in Eq.~\ref{eq:se1}, which assumes that $\rho^{NL}(m^2)$ vanishes as $m^2 \to \infty$ in the $m^2$ complex plane. Localizable densities $\rho^L(m^2)$ that grow polynomially also require renormalization; the standard procedure involves performing a finite number of subtractions until the UV divergences are isolated within the subtracted terms. In this section, we provide a brief overview of this procedure as a precursor to the nonlocal case, where an infinite number of subtractions are required.

For localizable spectral densities, the goal of renormalization is to isolate divergences within the first $N$ terms of the Taylor expansion of $\Sigma^0_h(p^2)$ about a regulating scale $p^2=m^2_0$, and subsequently subtract them:
\begin{eqnarray}
    \widetilde{\Sigma}_h(p^2;m_0^2)&=&
    \Sigma^0_h(p^2)-\Sigma^{0(N)}_h(p^2)+\nonumber \\
    &+&\left( \Sigma^{0(N)}_h(p^2) - 
    \sum_{n=0}^{N-1}\delta_n(p^2-m_0^2) \right),
\label{eq:ren1}
\end{eqnarray}
where the subtraction polynomial is defined as:
\begin{eqnarray}
\Sigma^{0(N)}_h(p^2)&=&\Sigma^{0}_h(m_0^2)+
\frac{d\Sigma^{0}_h}{dp^2}\Bigg\rvert_{p^2=m_0^2}(p^2-m_0^2)+\dots  \nonumber\\
&+&\frac{1}{(N-1)!}
\frac{d^{(N-1)}\Sigma^{0}_h}{dp^{2(N-1)}}\Bigg\rvert_{p^2=m_0^2}(p^2-m_0^2)^{N-1}. \nonumber
\end{eqnarray}
Since we possess the freedom to add a polynomial of contact terms with arbitrary coefficients $\delta_n$ to the two-point function, the $(N-1)$-degree polynomial $\Sigma^{0(N)}_h(p^2)$ can be canceled term-by-term. The resulting renormalized self-energy for a finite number of subtractions $N$ is:
\begin{eqnarray}
\widetilde{\Sigma}_h(p^2;m_0^2)
&=& \Sigma^0_h(p^2)-\Sigma^{0(N)}_h(p^2)\nonumber \\ 
&=&
-\int_{m^2_{th}}^{\infty} 
\frac{(p^2-m^2_{0})^N}{(m^2-m^2_0)^N}
\frac{\rho^{}_{\Sigma}\left(m^2\right)}{p^2-m^2+i\epsilon} 
dm^2 \nonumber \\
&=&
-\int_{m^2_{th}}^{\infty} 
\frac{g(p^2)}{g(m^2)}
\frac{\rho^{}_{\Sigma}\left(m^2\right)}{p^2-m^2+i\epsilon}dm^2, 
\label{eq:renN}
\end{eqnarray}
where $g(m^2) = (m^2-m^2_0)^N$ in the denominator acts as a damping function. This damping ensures that the theory is renormalizable if the spectral density $\rho(m^2)$ grows no faster than $(m^2)^N$. This effectively isolates the UV divergence within the first $N$ derivatives of the self-energy evaluated at the subtraction point $m_0^2$.

\subsection{\label{sec3:renormInf} Infinite-subtraction Renormalization}
In the case of a nonlocal spectral density, an infinite number of subtractions is required. To maintain consistency with the nonlocal form factors introduced previously, the natural choice for the damping function $g(p^2)$ in Eq.~\ref{eq:renN} is an entire function of the same form:
\begin{equation}
g(p^2) = \exp\left(\frac{p^2-m_0^2}{\Lambda_{NL}^2}\right).
\end{equation}
The renormalized nonlocal self-energy then takes the form:
\begin{eqnarray}
&& \nonumber \\
\widetilde{\Sigma}_h^{}(p^2)
&=& 
-\int_{m^2_{th}}^{\infty} 
\frac{g(p^2)}{g(m^2)}
\frac{\rho^{NL}_{\Sigma}\left(m^2\right)}{p^2-m^2+i\epsilon} 
dm^2\nonumber\\
&=& 
-\int_{m^2_{th}}^{\infty} 
\frac{
\exp\left(\frac{p^2-m_{0}^2}{\Lambda_{NL}^2}\right)
}
{
\exp\left( \frac{m^2-m_{0}^2}{\Lambda_{NL}^2}\right)
}
\frac{\rho^{NL}_{\Sigma}\left(m^2\right)}{p^2-m^2+i\epsilon} 
dm^2,\nonumber
\end{eqnarray}
The exponential in the numerator can be pulled outside the integral, which remains finite for any finite $p^2$. The resulting exponential increase of the self-energy at large timelike $p^2$ reflects an exponentially increasing Higgs decay amplitude. This is a fundamental physical consequence of our model: it provides the mechanism that damps all scattering amplitudes at large timelike momenta $p^2 \gg \Lambda^2_{NL}$.
By utilizing the expression for the exponentially increasing spectral density, 
\begin{equation}
    \rho^{NL}_{\Sigma}(m^2)=\exp\left(\frac{m^2-m_{th}^2}{\Lambda^2_{NL}}\right)\rho^{}_{1L}(m^2), \nonumber
\end{equation}
and setting the arbitrary subtraction scale to the threshold, $m_0^2 = m_{th}^2$, we arrive at the simplified expression:
\begin{eqnarray}
\widetilde{\Sigma}_h(p^2)
&=& 
-
\exp\left(\frac{p^2-m_{th}^2}{\Lambda_{NL}^2}\right)
\int_{m^2_{th}}^{\infty} 
\frac{\rho^{}_{1L}\left(m^2\right)}{p^2-m^2+i\epsilon} 
dm^2.\nonumber\\
\label{Eq:RNSE}
\end{eqnarray}
In this factorized form, the local 1-loop self-energy is modified by an exponential entire function acting as a form factor. Since the imaginary part of the 1-loop result is well-known, we obtain an analytical expression for the nonlocalizable self-energy:
\begin{eqnarray}
    \textrm{Im}[\widetilde{\Sigma}_h\left( p^2 \right)]
    &=& \exp\left(\frac{p^2-m_{th}^2}{\Lambda_{NL}^2}\right)
    \textrm{Im}[\Sigma_{h,1}\left( p^2 \right)]\nonumber\\
    &=&
    \frac{\lambda^2\Lambda_{NL}^2}{4\pi}
    \exp\left(\frac{p^2-m_{th}^2}{\Lambda_{NL}^2}\right)\sqrt{1-\frac{m_{th}^2}{p^2}},
    \label{eq:imS}\nonumber \\
\end{eqnarray}
where $\Lambda_{NL} \simeq m_{th}$ represents the scale of new physics. This scale acts as a UV regulator; beyond it, nonlocal effects become dominant and scattering amplitudes are exponentially suppressed. While $\Lambda_{NL}$ is technically arbitrary, in this framework it is naturally tied to the heavy-scalar threshold $m_{th}$, where multi-particle Higgs exchanges become kinematically accessible.

The emergence of a nonlocality scale just above $m_{th}$ is a direct consequence of the Higgs field vacuum condensation. The Higgs boson $h$ retains a "memory" of its coupling to the BSM sector. Below the threshold ($p^2 < m_{th}^2$), the heavy scalars decouple, leaving only the standard Higgs self-interactions with an effective quartic coupling $\lambda \simeq 0.1$. While the exact details of the SM Higgs potential emergence are outside the scope of this work, this decoupling ensures that the theory smoothly reproduces SM physics in the infrared.

\subsection{\label{Adelic} Renormalization as Adelic Completion}

In this section, we expand on the specific choice of the function $g(p^2)$ in the ratio $g(p^2)/g(m^2)$, which shares the same form as the nonlocal factor $\exp[(p^2-m^2_{th})/\Lambda^2]$. Given that infinite-derivative NLQFTs are often inspired by string theory, the unrenormalized K\"{a}ll\`{e}n-Lehmann amplitude can be viewed as the Archimedean (Real) completion of an underlying rational structure, $A_\infty(p^2)$. Following the construction of Adelic string amplitudes~\cite{Freund:1987ck, Brekke:1988ef}, the damping exponential $1/g(m^2)$ is identified as the product of all non-Archimedean ($p$-adic) completions: $\prod_{p} A_p(p^2)$. The global product,
\begin{equation}
\mathcal{A}(p^2) = A_\infty(p^2) \prod_{p} A_p(p^2),
\label{Eq:adele}
\end{equation}
defines the total Adelic amplitude. By the Adelic Product Formula, this global volume is a constant—which we uniquely identify with the renormalized physical amplitude.
The unrenormalized amplitude results from an infinite sum of local scalar interactions, manifesting as an infinite polynomial in $p^2$:
\begin{equation}
A_\infty(p^2) \sim \sum_{k=0}^{\infty} a_k \left(\frac{p^2}{\Lambda_{NL}^2}\right)^k \sim e^{n},
\label{Eq:rationals}
\end{equation}
where $n$ represents the number of Higgs-boson exchanges. Requiring all interactions to have unit couplings ensures that the interaction Lagrangian is defined by a formal power series over the field of rational numbers $\mathbb{Q}$. Consequently, the unrenormalized amplitude is a power series with rational coefficients, possessing a unique Archimedean completion and a corresponding set of $p$-adic completions. Crucially, according to Ostrowski's Theorem~\cite{Ostrowski1916, Koblitz1984}, these completions are the only possible non-equivalent valuations of $\mathbb{Q}$, and are fully determined by the transcendental structure of the Archimedean form factor.
In this Adelic framework, the standard operation of fixing subtraction-term coefficients order-by-order is replaced by an automatic subtraction across all orders. This infinite set of coefficients is effectively reduced to a single physical parameter: the nonlocality scale $\Lambda_{NL}$. The resulting amplitude is finite and vanishes in the ultraviolet limit.

\subsection{\label{Towers}The Higgs Resummed Propagator}
As established in Section~\ref{NLQFTSD}, the multiscalar interaction BSM theory presented here is a nonlocal scalar QFT. The Higgs kinetic term in the Action (Eq.~\ref{NLagrangian2}) yields the free nonlocal Higgs propagator:
\begin{equation}
    {G}_0(p^2)=
    \frac
    {
    \exp\left(\frac{p^2-m_{th}^2}{\Lambda_{NL}^2}\right)
    }
    {p^2-m_{b}^2+i\epsilon}.
\end{equation}
where $m_b$ denotes the bare Higgs mass. Following renormalization, this mass receives both SM and BSM contributions from the self-energy. We now calculate the resummed nonlocal renormalized Higgs propagator via Dyson resummation:
\begin{eqnarray}
\widetilde{G}(p^2)&=&G_0 
\left[ 1 + \widetilde{\Sigma}_{h}G_0 
+ (\widetilde{\Sigma}_{h}G_0)^2 
+ (\widetilde{\Sigma}_{h}G_0)^3
+ \dots
\right] \nonumber \\
&=& 
\frac{
\exp\left(\frac{p^2-m_{th}^2}{\Lambda_{NL}^2}\right)
}
{p^2-m_h^2-
i\exp\left(\frac{p^2-m_{th}^2}{\Lambda_{NL}^2}\right)
\textrm{Im}[\widetilde{\Sigma}_h(p^2)]+i\epsilon}
\nonumber \\
\end{eqnarray}

\begin{widetext}
\begin{eqnarray}
\widetilde{G}(p^2)=
\frac{
\exp\left(\frac{p^2-m_{th}^2}{\Lambda_{NL}^2}\right)
}
{p^2-m_h^2-
i\frac{\lambda^2\Lambda^2_{NL}}{4\pi}
\exp\left(\frac{2(p^2-m_{th}^2)}{\Lambda_{NL}
^2}\right)
\sqrt{1-\frac{m_{th}^2}{p^2}}
+i\epsilon},
\label{Eq:renormProp}
\end{eqnarray}
\end{widetext}
where $m_h^2 = m_b^2 + \textrm{Re}[\widetilde{\Sigma}_{h}(p^2=m_h^2)]$ is the renormalized Higgs mass. For virtual loop contributions, the full K\"{a}ll\`{e}n-Lehmann spectral distribution must be integrated over the space-like region ($p^2 < 0$). According to Eq.~\ref{Eq:RNSE}, radiative corrections are exponentially suppressed near the nonlocality scale. This leads to a naïve estimate for the SM radiative contributions to the Higgs mass:
\begin{equation}
\delta m^2_h \simeq \frac{\Lambda^2_{NL}}{16\pi^2}
\left[6\lambda_{SM}+\frac{1}{4}
(3g^2+g^{\prime 2})-6y^2_t
\right],
\label{Eq:hier}
\end{equation}
where $\lambda_{SM}$ is the SM quartic self-coupling, $g$ and $g^\prime$ are the gauge couplings, and $y_t$ is the top-Yukawa coupling. To address the hierarchy problem, Eq.~\ref{Eq:hier} suggests that the nonlocality scale—beyond which space-like contributions are suppressed—should reside in the few-TeV range.
Similarly, for large timelike scales ($p^2 \gg m_{th}^2$), the resummed propagator is also exponentially suppressed: $\widetilde{G}(p^2) \sim \exp(-p^2/\Lambda_{NL}^2)$. This renders the theory UV finite. These two asymptotic limits—suppression in the deep Euclidean (space-like) and deep Minkowski (time-like) regimes—make the emergent nonlocal theory a phenomenologically appealing framework.

\section{\label{PHENO}Higgs phenomenology at the High-Luminosity LHC}
LHC searches for heavy resonances and continuum excesses in di-photon, di-Higgs, and electroweak vector-boson final states \cite{CMS:2024nht,ATLAS:2023hbp,CMS:2022tqn,CMS:2018dqv,Li:2019pag,Hoffmann:2014aha} represent key efforts to explore new physics sectors beyond the Standard Model \cite{Arkani-Hamed:1998jmv,Randall:1999ee,Giudice:2016yja,Branco:2011iw}.
In the literature, it has been argued that the primary signature of nonlocality is a significant exponential modification of the measured cross section relative to SM expectations as the energy scale approaches the nonlocality scale, $p^2 \rightarrow \Lambda^2_{NL}$ \cite{BISWAS2015113,Su:2021qvm,Anderson:2024vzk}. As we have demonstrated for forward amplitudes, the renormalization of the two-point function leads to an eventual exponential suppression of the amplitude for $p^2 > \Lambda_{NL}^2$, a behavior also observed in recent studies of effective nonlocality \cite{Carone:2023cnp}. This transition can be summarized by the following ratio:
\[
\frac{\left|\mathcal{A}(X\rightarrow Y)\right|}
{\left|\mathcal{A}^{SM}(X\rightarrow Y)\right|} \simeq \begin{cases}
\exp\left(\frac{p^2-m^2_{th}}{\Lambda^2_{NL}}\right) & p^2\sim m^2_{th}\\
\exp \left(-\frac{p^2}{\Lambda^2_{NL}}\right) & p^2\gg m^2_{th}
\end{cases}\text{}
\]
A distinct signal of nonlocality at the LHC would be the production of multi-Higgs final states for $p^2 > m^2_{th}$. Final states involving two bosons (such as di-Higgs, di-photons, or electroweak di-bosons) provide clean experimental signatures, characterized by two back-to-back, ultra-high-energy objects in the LHC detectors. In this work, we evaluate the discovery potential of such di-boson analyses in the context of our proposed framework.

\subsection{\label{sec5:diHiggs} Vector Boson Fusion di-Higgs production}
As a primary application of our framework at the LHC, we propose measurements of $W^{\pm}, Z$, and Higgs-boson final states: $Vh, hh, VV$. When these are produced via Vector-Boson Fusion (VBF) or Vector-Boson Scattering (VBS) modes ($VV\rightarrow Vh, hh, VV$), the scattering amplitudes receive contributions from nonlocal effects, leading to deviations from the SM as described by Eq.~\ref{Eq:renormProp}.
\begin{figure}[hbt!]
\centering
\resizebox{0.47\textwidth}{!}{
\includegraphics{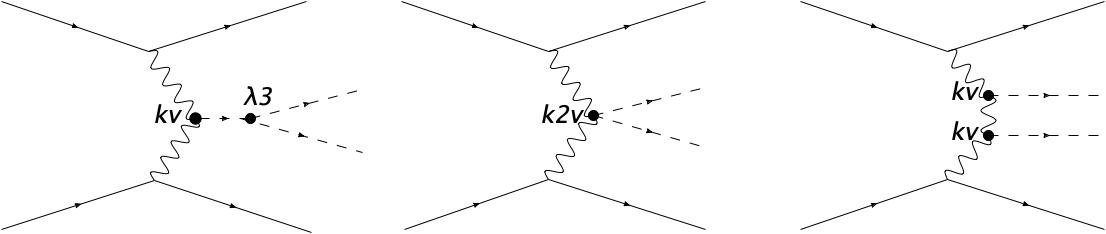}
}
\caption{Tree-level Feynman diagrams for VBF di-Higgs production: $s$-channel Higgs (left), quartic $VVhh$ channel (center), and $t$-channel $V$ exchange (right).}
\label{fig:VBFtoDiHiggs2}
\end{figure}
The leading-order (LO) di-Higgs VBF production diagrams are shown in Figure~\ref{fig:VBFtoDiHiggs2}. At high energy scales ($p^2 \gg M_W^2$), longitudinal vector boson scattering dominates the total amplitude \cite{Contino:2010rs}. In the absence of a BSM scale ($p^2 \ll \Lambda_{BSM}^2$), the LO amplitude is given by \cite{Bishara:2016kjn}:
\begin{equation}
\mathcal{A}(V_L V_L \rightarrow hh) \simeq \frac{s}{v^2}\left(k_{2V} - k_{V}^{2} \right),
\label{SMVVhhLO}
\end{equation}
where $s=p^2$ is the square of the center-of-mass energy. The modified couplings $k_{2V}$, $k_{V}$, and $\lambda_3$ (Fig.~\ref{fig:VBFtoDiHiggs2}) are normalized to their SM values:
\begin{equation}
k_{2V} = \frac{g_{2V}^{Meas}}{g_{2V}^{SM}}, \quad k_{V} = \frac{g_{V}^{Meas}}{g_{V}^{SM}}, \quad \lambda_3 = \frac{\lambda_{3}^{Meas}}{\lambda_{3}^{SM}}.
\label{vbfhhprod}
\end{equation}
In the high-$s$ limit of Eq.~\ref{SMVVhhLO}, the $s$-channel Higgs diagram is subleading and thus neglected, leaving as dominant the quartic $VVhh$ and $t$-channel $V$-exchange diagrams. In the SM ($k_{2V} = k_V = 1$), the amplitude remains small due to precise destructive interference between these graphs. Consequently, VBF di-Higgs production is notoriously difficult to measure at the LHC.
However, as $s$ approaches the nonlocality scale $\Lambda_{NL}^2$, the Higgs coupling to the BSM sector modifies the interaction. The amplitude then becomes (up to $\mathcal{O}(m_W^2/s)$ corrections):
\begin{equation}
\mathcal{A}(V_L V_L \rightarrow hh) \simeq \frac{s}{v^2} \left[ 
\widetilde{G}(p^2) - 1 \right].
\label{NLVVhhLO}
\end{equation}
Experimentally, VBF processes are uniquely "tagged" by two forward jets with a large rapidity gap. Under the nonlocal framework, the signal events should exhibit an excess with $s$ that depends on 
$\widetilde{G}(p^2)$. Such anomalous growth has been explored in composite Higgs models \cite{Bishara:2016kjn}, which similarly predict towers of states and exponential form factors.

Early signs of emergent nonlocality will manifest as a growth in the cross-section, which may initially resemble a high-mass resonance just beyond kinematic reach \cite{Carone:2023cnp}. However, the specific shape of this growth—determined by the spectral density—allows for a parameterized data analysis to constrain the nonlocality scale. Ultimately, $hh, Vh,$ and $VV$ final states serve as a unified probe of high-energy Higgs-field scattering, where longitudinal degrees of freedom dominate.

\subsection{\label{sec5:diPhoton} High-mass di-photon continuum searches}
The di-photon channel remains one of the cleanest final states at the LHC. Both the ATLAS and CMS collaborations have published extensive results on heavy scalar and tensor resonant and non-resonant production in di-photon final states \cite{CMS:2024nht,ATLAS:2023hbp,Wang:2021rvc,CMS:2018dqv}.

\begin{figure}[hbt!]
\centering
\resizebox{0.47\textwidth}{!}{
\includegraphics{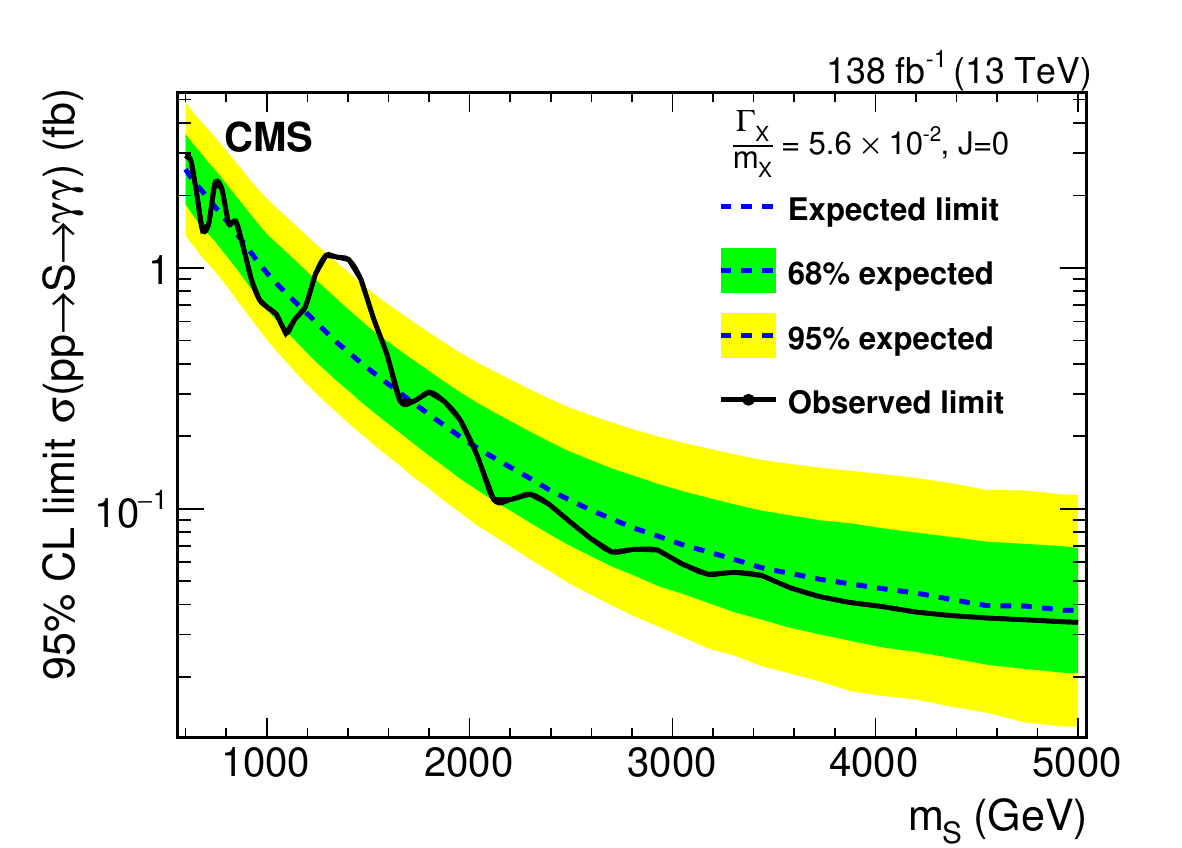}
}
\caption{Scalar di-photon resonance search at the LHC. Figure adapted from the CMS publication \cite{CMS:2024nht}.}
\label{fig:cmsgg}
\end{figure}

In this work, we utilize published results from the CMS experiment as a normalization baseline for future searches for nonlocalizable spectral densities. The CMS Run 2 search for a heavy scalar decaying to two photons is shown in Fig.~\ref{fig:cmsgg}, exploring a mass range from 0.5 to 5~TeV with an integrated luminosity of 138~fb$^{-1}$. The nonlocal BSM model proposed here predicts a specific continuum shape and differential cross-section $d\sigma/dM_{\gamma\gamma}$ which, as evidenced by Eq.~\ref{Eq:renormProp}, differs significantly from a standard Breit-Wigner resonance. While a new heavy state $\phi$ might appear just below the threshold and induce a threshold effect, we focus exclusively on characterizing the continuum behavior.

\begin{figure}[hbt!]
\centering
\resizebox{0.5\textwidth}{!}{
\includegraphics{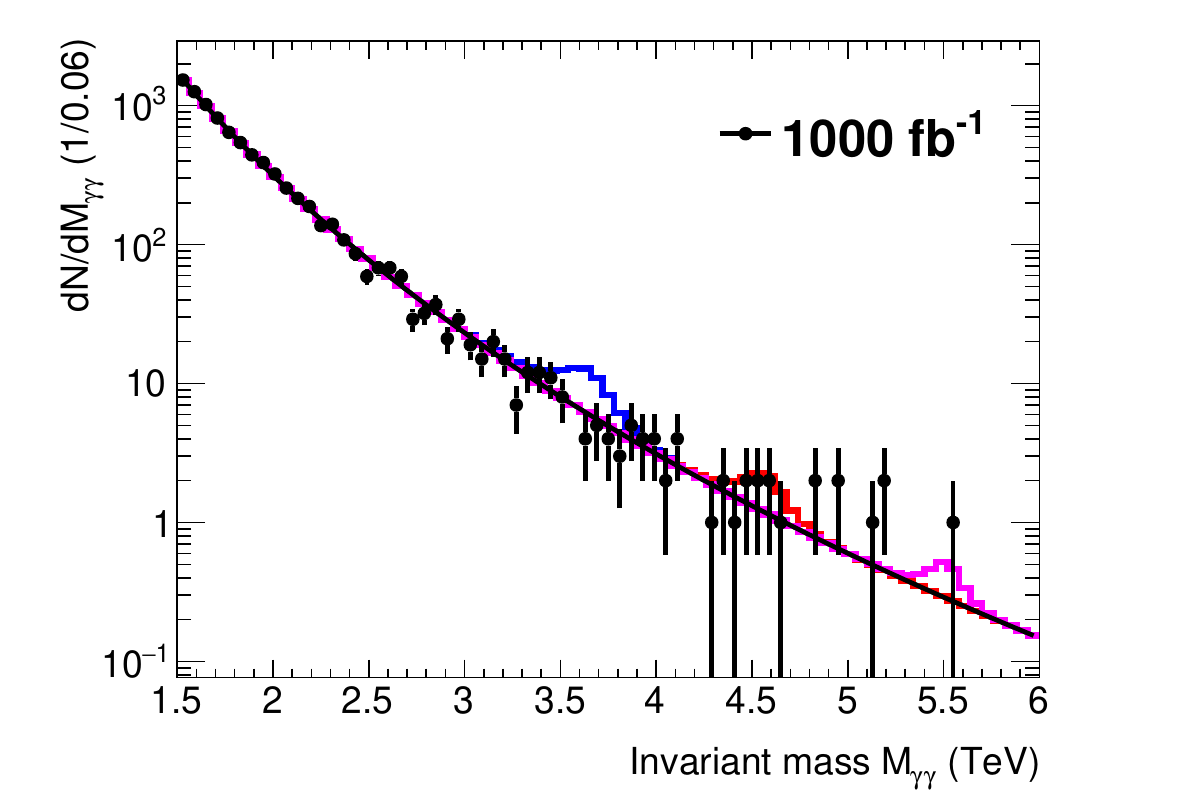}
}
\caption{Projection of the nonlocality di-photon continuum excess for an integrated LHC luminosity of 1000~fb$^{-1}$. The three peaks correspond to the three parameter sets discussed in the text. Randomly generated data points based on SM expectations are shown for comparison.}
\label{fig:cmsggSoB}
\end{figure}

The projected BSM nonlocal di-photon yields for three different parameter sets are shown in Fig.~\ref{fig:cmsggSoB}. We consider three nonlocality scales above a threshold of $m_{th}=3$~TeV: $m_{th}+5m_h$, $m_{th}+10m_h$, and $m_{th}+15m_h$. These results were produced using a fast simulation of the BSM signal, integrated with the published CMS di-photon invariant mass distributions for background normalization \cite{CMS:2024nht}. We applied the same event selection criteria as the CMS publication, assuming a di-photon efficiency of 65\% for masses above 1~TeV. While these signal predictions are not yet used to set formal limits, superimposing them onto realistic background distributions illustrates the unique signal shape (Fig.~\ref{fig:cmsggSoB}). Notably, the signal exhibits a relatively narrow mass distribution with a rising edge that is slower than its falling edge—a direct reflection of the nonlocal propagator where the imaginary part of the self-energy increases exponentially.

\begin{figure}[hbt!]
\centering
\resizebox{0.5\textwidth}{!}{
\includegraphics{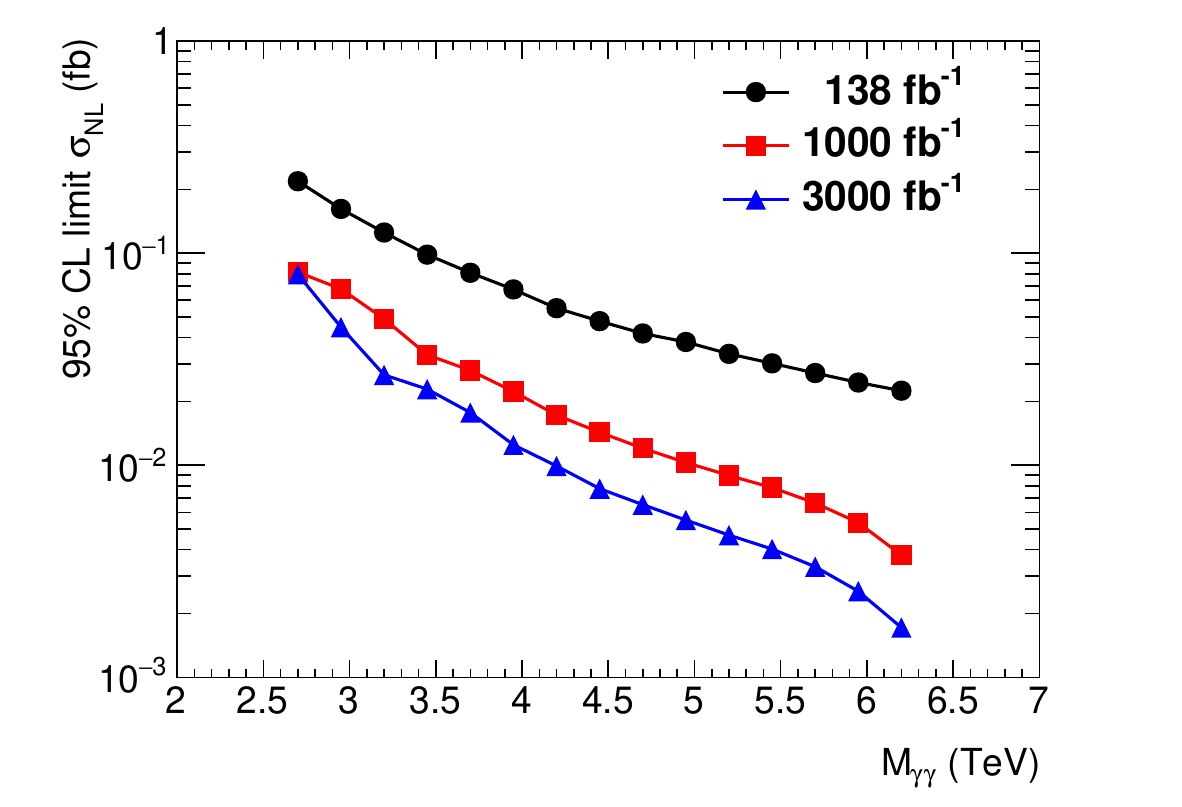}
}
\caption{Expected 95\% CL cross-section limits for BSM excesses with nonlocality scales up to 6.25~TeV across three integrated luminosities.}
\label{fig:cmsggLimit}
\end{figure}

We further explore the discovery potential of nonlocal Higgs spectral densities at the HL-LHC. In Fig.~\ref{fig:cmsggLimit}, the 95\% expected Confidence Level (CL) exclusion limits based on the background-only hypothesis are presented for three integrated luminosities \cite{Cowan:2010js,Paganis:2008xx}. The 138~fb$^{-1}$ projection reproduces the published Run-2 reach, while the higher luminosity projections represent the potential of the High-Luminosity LHC to achieve a 10-fold improvement in cross-section sensitivity.

\section{Discussion}
The Standard Model Higgs boson may be associated with an emergent nonlocalizable K\"{a}ll\`{e}n-Lehmann spectral density characterized by a nonlocality scale in the few-TeV range. For energies exceeding this scale ($p^2 > \Lambda_{NL}^2$), the scattering amplitudes are exponentially suppressed. The real part of the Higgs self-energy is suppressed at deep spacelike momenta ($p^2 \sim -\Lambda^2_{NL}$), effectively mitigating the Hierarchy problem. This emergent nonlocality arises from a Hamiltonian with an infinite number of local multiscalar interactions, where the exponential growth of the density of states $\rho(m^2)$ is driven by the rapidly increasing degeneracy of mass eigenstates.
The resulting framework is mathematically equivalent to an infinite-derivative nonlocal QFT. By assuming a nonlocal Higgs propagator and its corresponding K\"{a}ll\`{e}n-Lehmann representation, we have derived a renormalized self-energy by introducing a reciprocal form factor into the spectral integrand. The imaginary part of this self-energy, which constitutes the 1PI spectral density, was then used to obtain the full renormalized Higgs propagator and generate signal models for LHC searches.
Such nonlocal scalar sectors are accessible in current and future LHC runs. We argue that the nonlocal K\"{a}ll\`{e}n-Lehmann spectral density can be rigorously constrained through a simultaneous global fit of LHC measurements in exclusive channels, including di-Higgs, vector-boson, and di-photon production at high $\sqrt{s}$. Moving towards the direct extraction of real, positive-definite spectral densities $\rho(m^2)$ from collider data may represent a paradigm shift in the search for BSM physics.
This class of models shares significant overlap with the Higgsplosion program \cite{Khoze:2017ifq}, which involves non-perturbative multi-Higgs interactions. Although originally proposed as a strictly localizable theory, subsequent work has suggested that the Higgsplosion mechanism remains robust within a nonlocalizable QFT framework if the localizability requirement is relaxed \cite{Khoze:2018qhz}. 

In conclusion, testing general, non-perturbative spectral densities—incorporating bound states, resonances, and multi-particle continua—offers a powerful and inclusive approach to BSM discovery. The possibility of a nonlocalizable Higgs spectral density with a scale near a few TeV is particularly intriguing; it ensures finite scattering amplitudes in both deep spacelike and timelike regimes while providing a natural solution to the Hierarchy problem. As an emergent property of a light scalar sector with high degeneracies, such theories are compelling candidates for the fundamental origin of the SM Higgs potential.

\section*{Acknowledgements}           
This work was supported by a Taiwanese NSTC grant 
112-2112-M-002-020-MY3. 

\bibliography{NonLocalityHT}

\end{document}